# An Experimental Study on Shear and Flexural Strengthening of Concrete Beams Using GFRP Composites

Mohammad Rashidi[1,*], Hana Takhtfirouzeh[2]

[1]Department of Civil Engineering, University of Texas at El Paso, Texas, USA.
[1]Department of Civil Engineering, University of Memphis, El Paso, Tennessee, USA.

**Abstract:** Maintaining and restoring buildings has nowadays gained special importance due to its high cost. Various different methods have been presented regarding this matter due to the above-mentioned reason and also the ever-growing demand for the engineers and the building-industry specialists for strengthening, restoring, and improving the concrete structures. The effect of FRP sheets on shear and flexural strengthening of simple concrete beams has been examined in this research. Four concrete beams containing concrete were constructed with a similar cross section and length and a common resistance level. They underwent the two-point flexural test and their exploitation conditions were studied. The results showed that if the FRP sheets are used appropriately and with proper bracing in strengthening the concrete beam samples, there will be a significant increase in their shear and flexural strength. Moreover, it was observed that relative deformation measured on the concrete surface and the GFRP sheet is proportional.

## 1. Introduction

Vast researches have been conducted in the past few years based on using FRP (Fiber Reinforcement Polymers) sheets in order to strengthen reinforced concrete beams. The technology of using FRP sheets was first discovered in Swiss Federal Laboratory in 1984 [1]. FRP sheets have become widely popular in improving and restoring structures, especially concrete structures, in the past recent years since they are highly resistant to heavy loads, they are resistant to corrosion and chemicals, they are resistant to the fatigue caused by loading, and they can be quickly installed [2-5]. FRP sheets weigh 20% the weight of steel and are 2 to 10 times more resistant than steel [6]. The above-mentioned fibers are widely used in various industries for these superior properties [7]. FRP fibers have been used in the aviation industry for many years however their use in construction industry was limited since these fibers were relatively expensive in the past but a proper economical explanation can be presented for using them now because these materials are being massively produced today and so their prices have decreased. With regard to the fact that this strengthening technique is new, a lot of work has been done in the field of examining the behavior of these polymers in the flexural strengthening of concrete beams through sticking these fibers to the under- tension areas. All these studies have emphasized the mechanical-behavior improvement and increased flexural strength of the beams [8, 9].

In FRP-strengthened beams failure may occur due to beam shear, flexural compression, FRP rupture, FRP debonding or concrete cover ripping as presented by Ascione, et al [10], and Bonacci, et al [11, 12]. Based on experimental results conducted by Teng et al [13], the most common failure mode is due to debonding of FRP plate or ripping of the concrete cover. These failure modes are undesirable because the FRP plate cannot be fully utilized. In addition, such premature failures are generally associated with the reduction in deformability of the strengthened members. Premature failure modes are caused by interfacial shear and normal stress concentration at FRP cut off points and at flexural cracks along the beam. Extensive testing of such strengthened members has been carried out over the last two decades. A number of failure modes for RC beams bonded with FRP soffit plates have been observed in numerous experimental studies to date [14-33].

It is obvious that in addition to the resistance aspects, the performance of the under- exploitation members must be satisfactory in order to thoroughly examine the strengthened beams and this is not realized through merely providing the member with sufficient strength. The displacements caused under exploitation loads may be extremely large in a member which has been designed through ultimate strength method in such a manner that it damages the non-structural section. On the other hand, the initiated cracks in the beam may be so large that they could cause corrosion in the armatures which will also compromise the appearance.

This paper studies bond capacity in reinforced concrete beam and its influence on the flexural strength. Firstly, it introduces materials and test methods. Then it presents the comparison of the results of the experiment with the existing theories.

## 2. Materials and Methods

The effect of FRP sheets on the flexural strengthening of simple concrete- containing concrete beams with a common level of resistance has been examined in this experimental study. The manner of attaching the FRP sheets when constructing the samples and strengthening them have been considered as variables. Four concrete- containing concrete beams were constructed with the same cross section and length and a normal level of strength. They underwent a two- point flexural test and their exploitation conditions were examined. Among the four above- mentioned samples, one sample was constructed without FRP and was used as the control sample, two samples were constructed with one FRP layer underneath the beam for flexural strengthening, and one sample was built with two FRP layers on both sides for shear strengthening.

Four concrete- containing concrete beams were constructed in this research with the same cross section and length and a normal level of strength and they underwent a two- point flexural test until they experienced failure. The beams were divided into two groups. One beam was examined on without being strengthened and the rest of the samples were strengthened with one glass-fiber layer and then they were loaded. All the beams which were experimented on were 60 centimeters long they were placed on a 55-centimeter- wide support and were loaded and tested. The loading model of the beams can be seen in Figure 1. With regard to the results of the previous experiments [34], the length of the consumed FRP was considered to be equal to 65 centimeters in order to increase the effect of the strengthening and delay the separation of FRP from the concrete surface. This length would cover the entire web opening length of the beam plus parts of the flanges at the end of the beam.

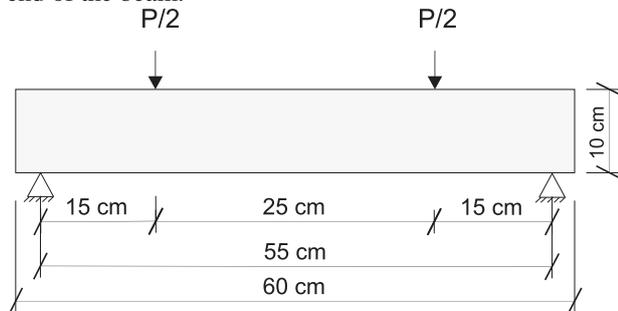

Fig. 1. The loading model of the beam

All the beams have 10×10 cm2 cross sections. Two concentrated symmetrical loads which were 25 centimeters away from each other were used for loading purposes. And so the shear span is equal to 25 centimeters and the ratio of the shear span length to the effective depth will be equal to 2.5 which places the intended beams in the class of common beams. The beams which have been experimented on have been shown in Figure 2.

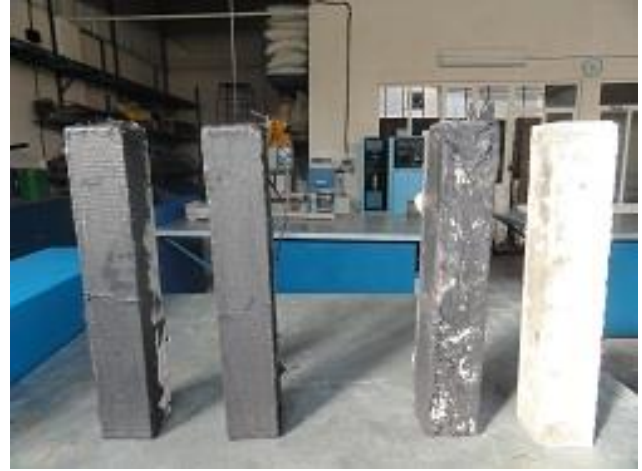

Fig. 2. The beams which have been experimented on

## 2.1 Material Properties

A 15×30 centimeter cylinder has been constructed in this experiment when concreting the samples and it was processed under conditions similar to that of beams. These samples underwent a pressure test when they were 28 days old their obtained compressive strength was equal to 34.5 MPa. The FRP used in this research was made of glass with a density equal to 1800 Kg/cm3. Each layer was 0.1 millimeters thick. This material's behavior was linear up to the failure point. The manufacturing company had announced its compressive strength and its elasticity module equal to 1100 and 35000 MPa respectively. The failure strain of the utilized FRP is equal to 1.7 percent. The combination plan of the utilized concrete is in accordance with Table 1.

G400 adhesive was used in this experiment. This adhesive is a two- component adhesive made by combining two adhesives, one as base and the other as the main adhesive substance which has been used to stick the samples with a proportion of 3 to 1 respectively, before combining them.

Table 1. Concrete combination plan

| Component | Weight ratio (kg/m³) |
|---|---|
| Cement | 500 |
| Gravel | 800 |
| Sand | 800 |
| Water | 220 |
| Sum | 2320 |

## 2.2 Sample Preparation

The properties of the used GFRP materials and the intended combination plan were presented in the previous section. The utilized concrete was first prepared in accordance with the combination plan by using the mixer in order to prepare the samples. 4 beam samples have been constructed along with a cylinder- shaped sample which is used to obtain the compressive strength of the concrete. The samples were kept in the water pool for a period of 28 days. The samples were brought out of the pool after 28 days and were kept in fresh air for one day when they were prepared for attaching the GFPR sheets. Based on the opinion of the researchers, it is advisable to keep the samples in fresh air for 3 to 7 days and then add the GFRP sheets. Anyhow, the GFRP sheets were cut as shown below after one day and were strengthened through shear and flexural methods by using the adhesive which was previously prepared by combining the two adhesives with a 3 to 1 proportion. The prepared samples were kept in fresh air for a day and then the test was conducted similarly in the laboratory for flexural and shear samples. Loading of the samples continues to the failure points. The manner of cutting and executing the FRP on the concrete samples are shown in Figure 3.

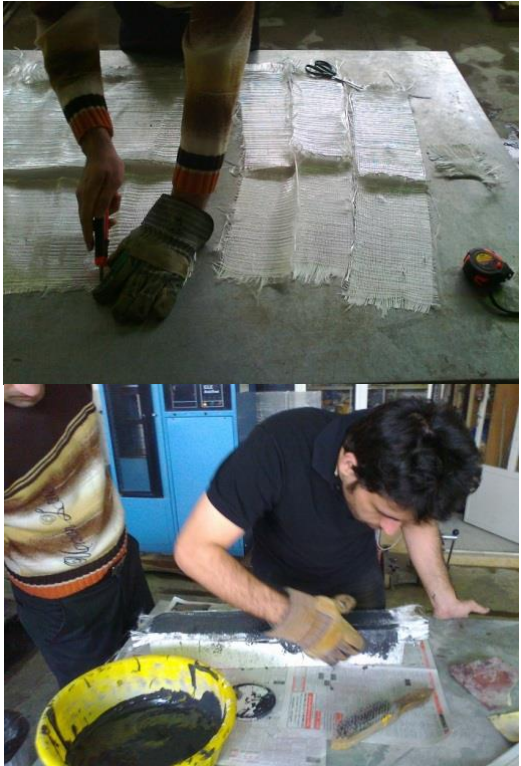

Fig. 3. Cutting and attaching the GFRP sheets

## 3. Results and Discussion

The beams under bending and also the changes made in the values of stress and strain in different stages of loading could be examined under the effect of concentrated loads. The force- displacement curve of these beams have three different gradients the first section of the curve is related to the cracking of tensional concrete district, the beam is very hard at this stage (like the beam which has not been strengthened) the second section is related to the cracking of tensional concrete district to the rebar yielding where the gradient of the graph decreases, however, it is still harder than the beam which has not been strengthened and the third section is related to the interval between the yielding point of the rebar and the point FRP ruptures or when FRP is separated from the concrete's surface. With regard to the relatively high elasticity module of FRP materials, attaching them on beams increases hardness and decreases deflection. Also, the FRP materials have a linear stress- strain behavior up to the rupturing point. Without absorbing sufficient energy and not having a yielding area like the yielding area of steel, it will rupture. Therefore installing them on the beam decreases ductility and energy absorption it is worth mentioning that the decrease in energy absorption is actually due to the local yielding of the longitudinal rebar in the rupture zone of FRP since the non-strengthened beams will have cracks in large areas and the rebar reaches the yield point, while in case of the strengthened beams, the rebar only changes shape in the rupture stage of FRP and it absorbs energy. In general, the ductility of the beam decreases as the FRP sheets increase. Therefore with regard to the fact that the value of the final strain of FRP materials is much larger in comparison with steel, when these materials are attached to the flexural surface of the beam through epoxy in order to increase the flexural strength of the beam, the steel reaches its yield point before the FRP composite starts tolerating a heavy load. Therefore it is not possible to increase the beam's hardness or its yield load without increasing the cross section of FRP, for more cooperation in transporting the beam, before the steel reaches its yield point. Strengthening the beam with the FRP system increases the final capacity of the cross section in general.

Shear strengthening of a concrete beam must also be considered if the concrete beam performs poorly in tolerating shear stress or when it tolerates the shear stresses less than the flexural capacity after flexural strengthening. Shear strengthening is in most cases a key and fundamental stage of effectively strengthening concrete beams. Using the bottom plane of FRP for flexural strengthening of reinforced concrete beams does not increase its shear strength that much; therefore, the flexural strengthening of the beams is not considered when planning the shear strengthening of the beams. It must be emphasized here that although the fibers attached to the lateral surfaces along the length of the beam do not help increase the beam's shear capacity, the fibers attached to these sides increase the shear strength of the beam from other angles. Except for the angles which are parallel to the shear cracks, most of the angles are effective in and beneficial for the purpose of attaching fibers to the sides in order to eliminate and decreasing the width of the cracks. Varied plans have been suggested for using FRP material for shear strengthening. These plans include attaching FRP to the sides of the beam, using the U- shaped jacket to cover the sides and the bottom surface of the beam or wrapping the cross section with FRP stripes and fibers.

FRP had a significant impact on both the flexural-strengthened beams and the shear- strengthened beams as expected in such a manner that it displayed its effects on increasing flexural and shear strength in both models. Four samples were studied in this experiment. A simple sample which was not strengthened in any way, two samples which underwent flexural strengthening through using a sheet of FRP under the beam and a sample which underwent shear strengthening through using two FRP sheets on both sides of the beam. It is worth mentioning that it has not been used any sort of armature in these samples. As expected, the brittle failure occurred in the distance between the two stands in the simple sample based on Figure 4. Two similar samples for flexural strengthening were prepared with one FRP sheet in the bottom of the beam in this experiment. One of these samples broke due to premature bond-failure of FRP sheet, almost like the simple sample, it only demonstrated a slight increase in flexural strength due to the FRP sheet according to Figure 5. The second flexural sample showed a fairly good resistance against the load which could be because of the strong bond between the concrete and the FRP sheet this sample's failure was due to shear weakness as shown in Figure 6. This sample experienced shear failure. The last sample of this test which had undergone shear strengthening had also a strong bond between the concrete and the FRP layers and showed a good resistance under the loads. The failure of this sample was due to flexural weakness as shown in Figure 7. This failure indicates the proper behavior of this beam. The respective Table 2 and Figure 8 contain the failure force of beam sample obtained from the test.

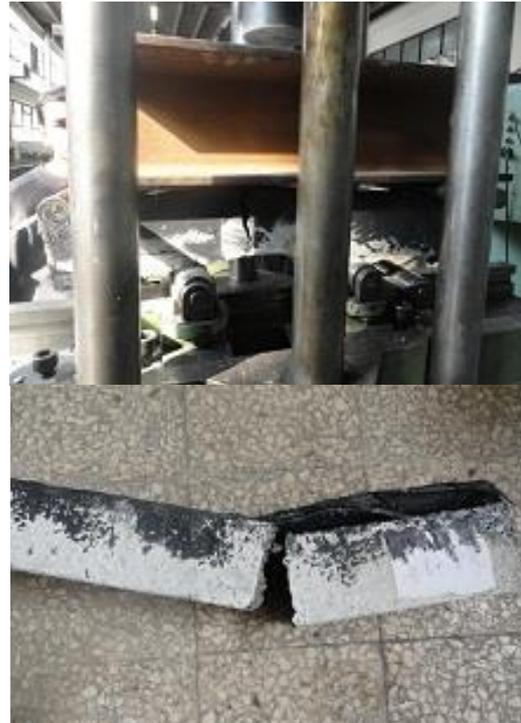

Fig. 6. Strong bond between the concrete and the FRP sheet

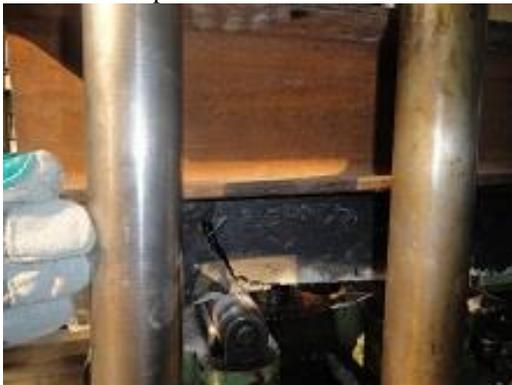

Fig. 4. Brittle failure in the distance between the two stands

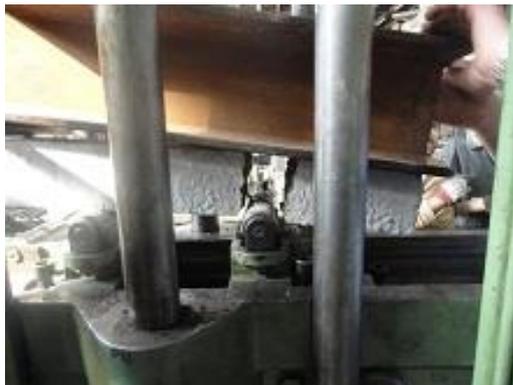

Fig. 5. Premature bond- failure of FRP sheet

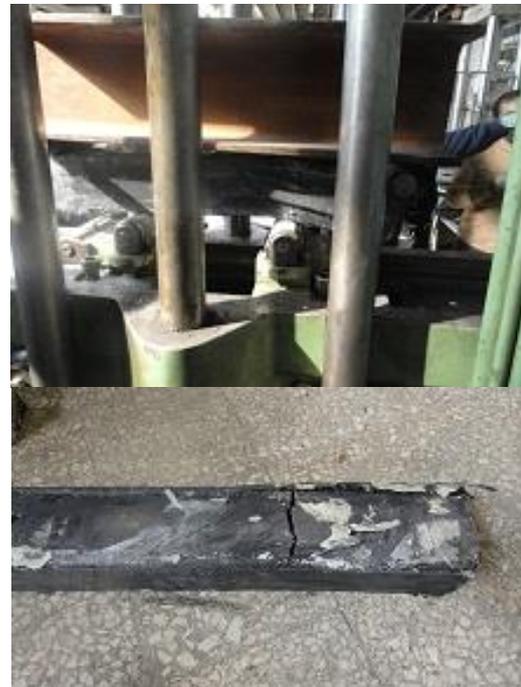

Fig. 7. Flexural weakness of the beams

Table 2. The results of the experiment

| Sample | Failure force (kg) |
|---|---|
| Simple beam | 817 |
| Beam with flexural layers 1 | 1097 |
| Beam with flexural layers 2 | 1437 |
| Beam with shear layers | 1117 |

Moreover, GFRP relative deformation has been measured to view the relative sliding between concrete and GFRP. This measure was conducted on the second flexural sample with flexural GFRP layers using manual and mechanical scales. As can be seen in Figure 9, relative deformation longitudinal GFRP sheets and the concrete surface at close GFRP sheets has been shown at the time of the failure loading. It can be concluded that relative deformation measured on the concrete surface and the sheet is proportional. The hypothesis of non-slip between concrete and sheet attached (cover sheet) has approved in this experiment.

## 4. Summary

This paper presents the application of GFRP in shear and flexural strengthening of concrete beams. The major conclusions derived from this experimental study are given as follows:

Strengthening the beam with the GFRP system increases the final capacity of the cross section in general.

Although the fibers attached to the lateral surfaces along the length of the beam do not help increase the beam's shear capacity, the fibers attached to these sides increase the shear strength of the beam from other angles.

The beam sample with shear layers of GFRP which had undergone shear strengthening had also a strong bond between the concrete and the FRP layers and showed a good resistance under the loads and the failure of this sample was due to flexural weakness.

Relative deformation longitudinal GFRP sheets and the concrete surface at close GFRP sheets has been presented at the time of the failure loading. The hypothesis of non-slip between concrete and sheet attached (cover sheet) has approved in this experiment.

It could eventually be stated that if the FRP sheets are used appropriately and with proper bracing in strengthening the concrete beam samples, there will be a significant increase in their shear and flexural strength. It could be claimed based on the results of researches [30] that FRP has a significant effect on increasing the ductility of concrete beams. This matter was not examined in the present research and the main objective of this study is to show the effect of FRP on increasing shear and flexural strength.

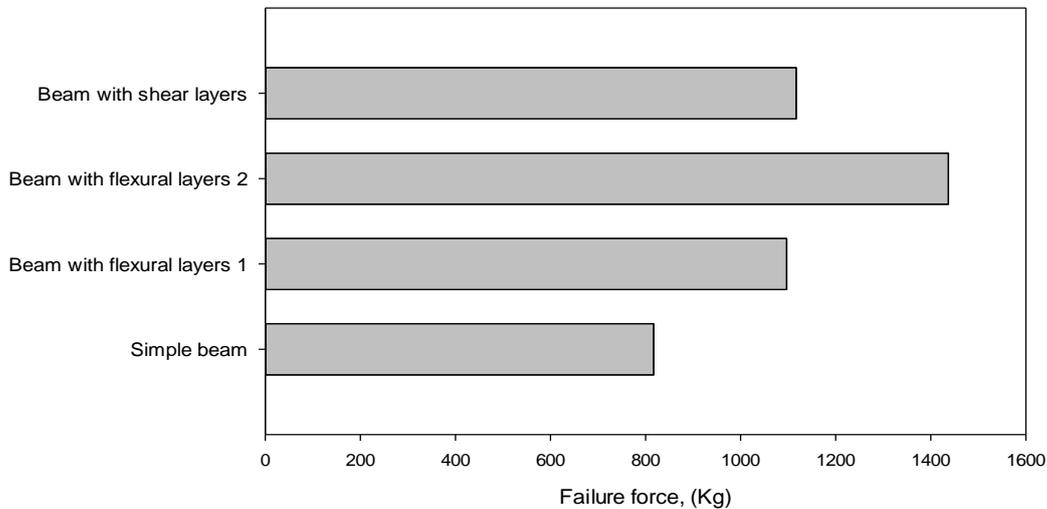

Fig. 8. Relative deformation longitudinal GFRP sheets and the concrete surface

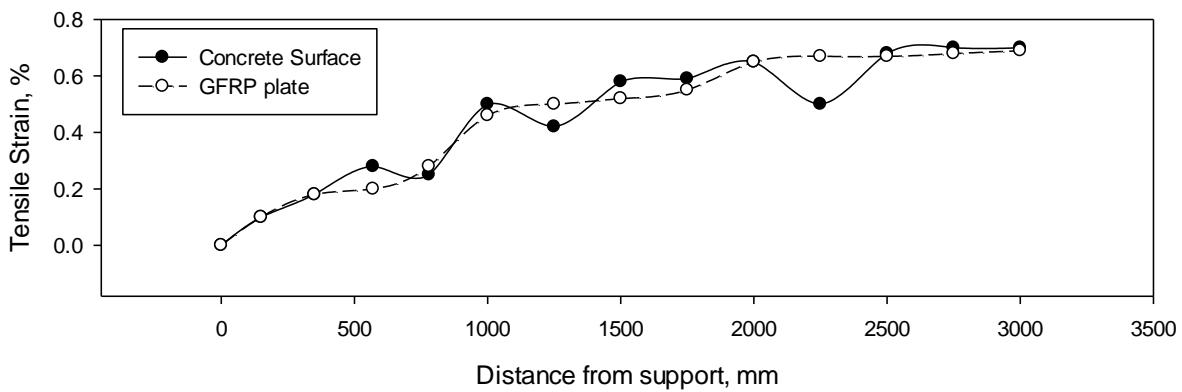

Fig. 9. Relative deformation longitudinal GFRP sheets and the concrete surface

*Corresponding author mrashidi@miners.utep.edu